\documentstyle[multicol,epsf,aps]{revtex}
\newtheorem{theorem}{Theorem}[section]
\newtheorem{lemma}{Lemma}[section]
\newcommand{\bm}[1]{\mbox{\boldmath $#1$}}
\def\p{{\bf p}}
\def\A{{\bf A}}

\def\R{{\bf R}}

\def\HS{{\rm HS}}

\def\E{{\cal E}}

\def\mfr#1/#2{\hbox{${{#1} \over {#2}}$}}
\def\uprho{\raise1pt\hbox{$\rho$}}
\def\upchi{\raise1pt\hbox{$\chi$}}
\def\dlambda{\lower1pt\hbox{$\lambda$}}
\catcode`@=11
\def\eqalignii#1{\,\vcenter{\openup1\jot \m@th
\ialign{\strut\hfil$\displaystyle{##}$&
        $\displaystyle{{}##}$\hfil&
        $\displaystyle{{}##}$\hfil\crcr#1\crcr}}\,}
\catcode`@=12

\def\be{\begin{equation}}
\def\ee{\end{equation}}
\title{
 Density Matrix Functional Calculations\\for Matter in Strong Magnetic 
Fields: 
\\I.\ Atomic Properties} % 
\author{Kristinn Johnsen$^{a,b,c}$\hspace{0.1cm} and \hspace{0.1cm}Jakob 
Yngvason$^{b,c}$
\\[0.5cm]${}^a$Mikroelektronik Centret, 
Technical University of Denmark, Bldg. 345e , DK 2800 Lyngby, 
Denmark\\[0.2cm]
${}^b$Science Institute, University of Iceland, Dunhaga 3, IS 107 
Reykjavik, Iceland
\\[0.2cm]
${}^c$NORDITA, Blegdamsvej 17, DK 2100 K\o benhavn \O , Denmark\\[0.1cm]}
\begin{document}
\maketitle
\begin{abstract} We report on a numerical study of the density matrix 
functional introduced by Lieb, Solovej and Yngvason for the investigation 
of heavy atoms in high magnetic fields. This functional describes {\em 
exactly} the quantum mechanical ground state of atoms and ions in the 
limit when the nuclear charge $Z$ and the electron number $N$ tend to 
infinity with $N/Z$ fixed, and the magnetic field $B$ tends to infinity in 
such a way that $B/Z^{4/3}\to\infty$.
We have calculated electronic density profiles and ground state energies 
for values of the parameters that prevail on neutron star surfaces and 
compared them with results obtained by other methods. For iron at 
$B=10^{12}$ G the ground state energy differs by less than 2 \% from the 
Hartree-Fock value. 
We have also studied the maximal negative ionization of heavy atoms in 
this model at various field strengths. In contrast to Thomas-Fermi type 
theories atoms can bind excess negative charge in the density matrix 
model. For iron
at $B=10^{12}$ G the maximal excess charge in this model 
corresponds to about one 
electron.
\end{abstract}
\begin{multicols}{2}
\narrowtext
\section{Introduction}

The properties of matter in magnetic fields of the extreme strength of 
$10^{12}$
Gauss and higher have been the subject of numerous investigations since 
the early seventies, a major impetus being the discovery of pulsars in 
1968 and the resulting interest in magnetized neutron stars. We refer to 
\cite{C92}, \cite{M92}, \cite{R94}, \cite{LSY94a},\cite{LSY94b} for general reviews on 
this subject and lists of references.  The standard Hamiltonian of atomic 
physics,
\begin{eqnarray}
H_{N,B,Z}&=&\sum_{i=1}^N\left\{ 
[(\p^{(i)}+\A(\bm r^{(i)}))\cdot{\sigma}^{(i)}]^2 -Z\vert 
\bm r^{(i)}\vert^{-1}\right\}\nonumber \\
&& + \sum_{1\leq i<j\leq N}\vert 
\bm r^{(i)}-\bm r^{(j)}\vert^{-1}\label{dm1} 
\end{eqnarray}
is usually taken as a starting point for the study of atoms in the 
atmosphere and outermost crust of neutron stars. Here $N$ is the number of 
electrons that move in the Coulomb field of a 
nucleus, localized at the origin with charge $Ze$, and in a homogeneous 
magnetic field ${\bf B}=(0,0,B)$ with vector potential $\A({\bm 
r})=(1/2)(-yB,xB,0)$.
The Hamiltonian (\ref{dm1}) operates on antisymmetric wave functions 
$\Psi\in\bigwedge_1^N L^2({\bf R}^3;{\bf C}^2)$ of space and spin 
variables, and 
$\sigma=(\sigma_1,\sigma_2,\sigma_3)$ is the vector of Pauli matrices. 
Units are chosen such that $\hbar=e=2m_e=1$, $c=1/\alpha\approx 137$; the 
energy unit is then four times the Rydberg energy, i.e., 54.4 eV. 
Besides the atomic Hamiltonian (\ref{dm1}) it is, of course, important to 
study the Hamiltonian for molecules and matter in bulk, but the present 
paper is only concerned with (\ref{dm1}), more specifically with its 
ground state energy 
\be
E^{\mbox{\tiny
Q}}(N,B,Z)=\inf_{(\Psi,\Psi)=1}(\Psi,H_{N,B,Z}\Psi),\label{dm2}\ee and the 
ground state 
electron density is
\begin{eqnarray}
&&\uprho^{\mbox{\tiny
Q}}_{N,B,Z}(x)=
N \sum_{s^{(i)}=\pm\mfr1/2}
\\ &&\times
\int\left|\Psi_0(x,x^{(2)},\ldots,x^{(N)} 
;s^{(1)},\ldots,s^{(N)})\right|^2 dx^{(2)}\cdots dx^{(N)}\label{dm3}
\nonumber
\end{eqnarray}
where $\Psi_0$ 
is a ground state wave function. 

Previous works on matter in strong magnetic fields can roughly be 
divided into two classes. On the one hand the focus has been on light 
atoms, in particular hydrogen with $Z=1$, on the other hand on heavy atoms 
with high $Z$. The present contribution falls into the second class. 
Here $Z=26$ plays a special role because iron is believed to be the most
abundant element in the surface layer of a neutron star \cite{C92}, \cite{M92}. 
For such heavy atoms it is reasonable to expect that important aspects can 
be extracted from an asymptotic analysis in $Z$, and since $10^{12}$ G
is large even compared with the natural atomic unit 
$B_0=m^2e^3c/\hbar^3=2.35\times 10^9$ G, an asymptotic analysis in $B$ is 
equally called for.\footnote{With our choice of units $2m_e=1$ and the 
magnetic field is actually measured in units of $4B_0=9.40\times 10^9$ G.}

The asymptotic behavior of 
the energy (\ref{dm2}) and density (\ref{dm3}) as $N,Z\to\infty$, where 
$N/Z$ is fixed 
and the magnetic field $B$ is allowed to vary with $Z$ as well, has 
recently been rigorously studied by Lieb, Solovej and Yngvason 
\cite{LSY92},  \cite{LSY94a}, \cite{LSY94b}. In these papers it was proved
that the ground state properties of (1) can in this limit be evaluated 
{\em exactly} by five nonlinear functionals corresponding to different 
physics at different scales of the magnetic field $B$ as measured by 
powers of $Z$. These five parameter regions are characterized as follows: 
{\em Region 1}, $B\ll Z^{4/3}$; {\em Region 2}, $B\sim Z^{4/3}$; {\em 
Region 3}, $Z^{4/3}\ll B\ll Z^3$; {\em Region 4}, $B\sim Z^3$; {\em Region 
5}, $B\gg Z^3$. Here $B\ll Z^p$, $B\gg Z^p$ and $B\sim Z^p$ means 
respectively that the ratio $B/Z^p$ tends to $0$, $\infty$ or a constant 
$\neq 0$ as as $Z\to\infty$.

The asymptotic theories corresponding to Regions 1-3 are semiclassical 
theories of Thomas-Fermi type that have been extensively applied to 
neutron stars in the past, see, e.g., 
\cite{K70},\cite{MRS71},\cite{C74},\cite{FGP89},\cite{AS91},\cite{RFGPY93}
and \cite{SO84}. 
Salient features of atoms in region 5 were captured by a different density 
functional theory already in the papers \cite{KK71b} and \cite{KK71a}. 
However, the conditions on the surface of a typical neutron star 
correspond rather to region 4, and this asymptotic region is also the most 
interesting one from the mathematical point of view. In fact, in 
\cite{LSY94a} it is shown that it can be described by a functional of a 
novel type, where the variable is not a density, but a function with 
values in density {\em matrices}. Moreover, this theory covers regions 3 
and 5 as limiting cases. We refer to it as the density matrix (DM) 
theory.

In view of the fact that the DM theory is an exact limit of quantum 
mechanics it is important to know its properties in some detail. Being an 
asymptotic theory it is clear that it does not encompass the same 
information as the full Hamiltonian at finite $Z$ and $B$. In particular 
the DM theory does not capture exchange-correlation effects, and it is a 
theory of very strong fields in the sense that all electrons are confined 
to the lowest Landau band. These features should not be considered as a 
shortcoming of the DM theory, however. In fact, the hardest part of the 
derivation of the limit theorems in \cite {LSY94a} is precisely to prove 
rigorously that contributions from exchange and higher Landau bands vanish 
in the limit considered. The DM theory should be judged in its own merits: 
It is enormously more simple numerically than the full quantum mechanical 
problem (\ref{dm2}) and it is a well defined starting point for more 
refined approximations.

In the present contribution we report on a numerical study of the DM 
theory for atoms. We have computed ground state energies and electronic 
density profiles
over a wide range of parameters and compared them with results obtained by 
different methods. In particular we compare the DM theory to the 
semiclassical theory that applies in Region 3, the simple density 
functional theory for Region 5, and also to other density functional 
\cite{J85}, \cite{KM88} and HF \cite{KOONIN87} calculations. The 
difference between DM and HF calculations of ground state energies is less 
than 2\% where data are available so that comparison can be made. This is 
remarkable in view of the fact that for standard Thomas Fermi theory with 
$B=0$ the Scott term, which corrects for the rough treatment of the 
electrons close to the nucleus in TF theory, must be incorporated in order 
to achieve such a good numerical agreement, cf.\ 
\cite{E88}. Thus, at least at this field strength,  DM theory is closer to 
HF theory than might have been expected. A more precise statement requires 
an analysis of the next to leading order terms in the asymptotic expansion 
of the ground state energy. Such an analysis has yet to be carried out.

Another point where the DM theory differs from semiclassical theories is 
in 
the possibility of negative ionization. It is a general feature of 
Thomas-Fermi type theories, based on potential theoretical arguments, cf.\ 
\cite{L82}, that the number of bound electrons never exceeds $Z$. For the 
quantum mechanical problem the meaning of this is simply that the binding 
energy of an excess electron must necessarily be of lower order in $Z$ 
than the ground state energy.  On the other hand it is known that a 
magnetic field enhances binding, for instance the 
Hamiltonian (1) with $N=Z+1$ has infinitely many bound states for $B\neq 
0$ \cite{AHS81}. In the limit of extremely strong fields in Region 5 the 
negative charge can even be as large as $2Z$ \cite{LSY94a}. 
The only rigorous results on the DM theory concern this extreme limit,
but our numerical computations clearly show negative ionization that
increases with $B$. It 
seems, however, that in order to approach the $2Z$ value extremely strong 
fields are needed; even at fields as strong\footnote{The computations at 
these extreme field strength were carried out mainly to test the 
mathematical properties of DM theory. It is clear that doubts about the 
applicability of the nonrelativistic Hamiltonian (\ref{dm1}) can be raised 
in such extreme fields, even for very heavy atoms.} as $10^{18}$ G the 
excess charge for iron is \lq\lq only\rq\rq\ about 23 \%.

Our interest in negative ionization is also motivated by its relation to 
another question, the binding of atoms into molecules and chains. Although 
a rigorous mathematical theorem linking these two aspects of binding does 
not seem to exist, it is a fact that in regions 1-3, i.e., for $B\ll Z^3$, 
molecular binding energies are vanishingly small compared to ground state 
energies, whereas in region 5 binding becomes extremely strong: For a 
diatomic molecule the binding energy is 6 times the ground state energy of 
an individual atom! The question whether iron is weakly or strongly bound 
at field strengths of the order
$10^{12}$ G has been controversial over the past 25 years. The best HF 
results \cite{KOONIN87} indicate weak or no binding, but the computations 
are difficult for they amount to subtracting one large number from 
another.  It is decisive to treat the molecules and the individual atoms 
consistently by the same numerical methods so that unavoidable errors 
cancel as far as possible. Since the DM theory is numerically much simpler 
than HF theory it is easier to achieve this in the former and we shall 
return to the binding question in a separate paper.
The atomic computations presented here are a necessary preparation for the 
study of molecules and chains.

\section{The Density Matrix Theory and its Limiting Cases}

The density matrix theory \cite{LSY92}, \cite{LSY94a} is based on an energy functional, where the 
variable is a mapping $\Gamma$: ${\bf r}_\perp\to\Gamma_{{\bf r}_\perp}$ 
from
${\bf r}_\perp=(x,y)\in\R^2$ into density matrices, i.e., nonnegative 
trace 
class operators on $L^2(\R,dz)$. In a magnetic field of strength $B$ these 
operators have to satisfy the condition \be 
0\leq\Gamma_{{\bf r}_\perp}\leq (B/(2\pi))I\label{dm4}\ee for all ${\bf 
r}_\perp$. 
Let 
$\Gamma_{{\bf r}_\perp}(z,z')$ denote the integral kernel of 
$\Gamma_{{\bf r}_\perp}$ and put $\uprho_\Gamma(\bm r)=\Gamma_{{\bf 
r}_\perp}(z,z)$ for $\bm r=(\bm r_\perp,z)\in\R^3$. 
The {\em density matrix functional} for an atom with nuclear charge $Z$ is 
defined by
\begin{eqnarray}
{\cal E}^{\mbox{\tiny DM}}[\Gamma ]
&=& -\int 
\left[
{\partial^2 \Gamma_{{\bf r}_\perp}(z,z')\over\partial z'^2}
\right]_{z'=z}
\, d^3\bm r - Z\int {\uprho_\Gamma (\bm r)\over |\bm r|}\,d^3\bm r  
\nonumber \\&&
+{1\over 2}\int\int {\uprho_\Gamma (\bm r)\uprho_\Gamma (\bm r')\over |\bm 
r-\bm r'|}
\,d^3\bm r'\,d^3\bm r.
\label{dm:n_1}
\end{eqnarray}

In the density matrix theory the electrostatic interactions are treated 
classically but the kinetic energy for the motion along the magnetic field 
quantum mechanically by the $-\partial^2/\partial z^2$ term. In directions 
perpendicular 
to the field the motion is restricted by the \lq\lq hard core\rq\rq\ 
condition (\ref{dm4}). This condition reflects the fact that the density 
of states per unit area for free electrons in the lowest Landau band is 
$B/(2\pi)$. The functional (\ref{dm:n_1}) is plausible if one thinks of 
$\Gamma_{{\bf r}_\perp}(z,z')$ as an approximation to
\begin{equation}
N\int\Psi_0({\bf r}_\perp,z;{\bf r}_2,\dots,{\bf r}_N)
\Psi_0^*({\bf r}_\perp,z';{\bf r}_2,\dots,{\bf r}_N)\prod_{j=2}^N d{\bf r}_j,
\end{equation}
where $\Psi_0$ is a normalized ground state wave function. In the parameter region $B\gg Z^{4/3}$ the electrons are confined to the lowest Landau band, and the Pauli Hamiltonian $[(\p+\A(\bm r))\cdot{\sigma}]^2$, restricted to the lowest Landau band, is precisely $-\partial^2/\partial z^2$.

The ground state energy for $N$ electrons in DM theory is 
\be E^{{\mbox{\tiny DM}}}(N,B,Z)=\inf\{{\cal E}^{{\mbox{\tiny 
DM}}}[\Gamma]: 
\hbox{$\int \uprho_\Gamma(\bm r)\,d^3\bm r\leq N$}\}.\label{dm6}\ee 
As shown in \cite{LSY94a}, Theorem 4.3, there is a unique minimizer for 
this variational problem, i.e.,
\be E^{{\mbox{\tiny DM}}}(N,B,Z)={\cal E}^{{\mbox{\tiny 
DM}}}[\Gamma^{{\mbox{\tiny DM}}}_{N,B,Z}]\ee
with a unique $\Gamma^{{\mbox{\tiny DM}}}_{N,B,Z}$. The corresponding 
density, $\uprho^{\mbox{\tiny DM}}_{N,B,Z}$, satisfies $\int 
\uprho^{\mbox{\tiny DM}}_{N,B,Z}=N$, if $N\leq N_c$, and
$\int \uprho^{\mbox{\tiny DM}}_{N,B,Z}=N_c$, if $N>N_c$,
where $N_c\geq Z$  is 
a number depending on $Z$ and $B$. As explained in the next section, the 
minimization problem (\ref{dm6}) amounts to seeking at each $\bm r_\perp$ 
the lowest eigenvalues and eigenfunctions for a one-dimensional 
Schr\"odinger Hamiltonian $-\partial^2/\partial z+V^{\mbox{\tiny DM}}_{\bm 
r_\perp}(z)$ where $V^{\mbox{\tiny DM}}_{\bm r_\perp}$ is the 
self-consistent potential generated by the nucleus and 
$\uprho^{\mbox{\tiny DM}}_{N,B,Z}$.

The density matrix theory is in fact a two parameter theory with 
parameters $\lambda=N/Z$ and $\eta=B/ Z^3$ due to the {\em scaling 
relations} 
\be E^{{\mbox{\tiny DM}}}(N,B,Z)=Z^3 E^{{\mbox{\tiny 
DM}}}(\lambda,\eta,1)\label{dm8}\ee 
and 
\be\uprho^{\mbox{\tiny DM}}_{N,B,Z}(\bm r)=Z^4\uprho^{\mbox{\tiny 
DM}}_{\lambda,\eta,1}(Z\bm r).\label{dm9}\ee 
In particular, the ratio to $Z$ of the maximal number of electrons that a 
nucleus can bind in DM theory, $\lambda_c=N_c/Z$, is a function of $\eta$ 
alone.

The DM theory holds a special position in the study of the properties of 
matter in strong magnetic field because it provides an {\em asymptotically 
exact} description of the quantum mechanical ground state energy 
$E^{\mbox{\tiny Q}}$ and electron density $\uprho^{\mbox{\tiny Q}}$ as 
$N$, $Z$ and $B$
tend to infinity with $N/Z$ fixed and $B/Z^{4/3}\to \infty$.
The following theorems are proved in \cite{LSY94a}, Theorems 1.1 and 8.1:
\begin{theorem}
Let $N$, $Z\to\infty$ with $N/Z$ fixed. If $B/Z^{4/3}\to \infty$, then 
\be E^{\mbox{\tiny Q}}(N,B,Z)/E^{\mbox{\tiny DM}}(N,B,Z)\to 1.\ee
\end{theorem}

\begin{theorem}
Let $N$, $Z$ and $B\to\infty$ with $N/Z=\lambda$ and $B/Z^3=\eta$ fixed.
Then 
\be Z^{-4}\uprho^{\mbox{\tiny Q}}_{N,B,Z}(Z^{-1}x)\to\uprho^{\mbox{\tiny 
DM}}_{\lambda,\eta,1}(x)
\ee
in the sense of convergence of distributions.
\end{theorem}

The shape of atoms in DM theory is discussed in Section IV in connection 
with Figs.\ 1 and 2. It should be kept in mind that by the limit theorems 
II.1 and II.2 the DM theory is a theory of {\em heavy} atoms. We have 
chosen iron with $Z=26$ as our reference because of its astrophysical 
importance. By the scaling relations (\ref{dm8}) and (\ref{dm9}) it is 
simple to transform the results to other values of $Z$. As seen from the 
figures the atom is approximately spherical when the magnetic field is not 
too strong ($<$ ca. $10^{11}$ Gauss for iron), but becomes increasingly 
elongated as the field goes up. In fact, as shown in \cite{LSY94a} the 
limiting cases $\eta\to 0$ and $\eta\to\infty$ of the DM theory can be 
described by simpler theories that we now review briefly, referring to
\cite{LSY94a} and \cite{LSY94b} for details. 

The weak field limit, $\eta\to 0$, is the Thomas-Fermi theory for atoms in 
strong magnetic fields, where only the lowest Landau band is taken into 
account
(as in DM theory). This theory was introduced by Kadomtsev \cite{K70} and 
studied further in a number of publications, see \cite{LSY94b} for a list 
of references. In \cite{LSY94a}, \cite{LSY94b} it is called the STF 
theory. The density functional is
\begin{eqnarray}
\E^{\mbox {\tiny STF}}[\uprho]&=&
{4\pi^4\over 3\, B^2}\int \uprho(\bm r)^3\,d^3 \bm r-
Z\int{\uprho(\bm r)\over \vert \bm r\vert}d^3\bm r\nonumber \\ 
&&+{1\over 2}\int\int 
{\uprho(\bm r)\uprho(\bm r')\over\vert\bm r-\bm r'\vert}\,d^3\bm r d^3\bm 
r'.\label{dm12}
\end{eqnarray}
The precise connection between DM and STF theory is given in 
\cite{LSY94a}, Eq.\ (8.11); if $E^{\mbox{\tiny STF}}(N,B,Z)$ denotes the 
infimum of (\ref{dm12}) with subsidiary condition $\int\uprho\leq N$, then
\be \lim_{\eta\to 0}E^{\mbox{\tiny DM}}(\lambda,\eta,1)/\eta^{2/5}=
E^{\mbox{\tiny STF}}(\lambda,1,1).\ee
In STF theory, atoms are {\em spherical} with a finite radius $\sim
Z^{-1/3}(B/Z^{4/3})^{-2/5}$.

In the opposite parameter regime, more precisely for $\eta$ larger than a 
certain critical value, $\eta_c$,   
DM theory also reduces to a density functional theory. The value of 
$\eta_c$ depends on $\lambda$; for $\lambda=1$ we find $\eta_c=0.148$, 
which for $Z=26$ corresponds to $B=2.44\times 10^{13}$ G. The energy 
functional appropriate for such {\em super strong} (SS) fields is
\begin{eqnarray}
\E^{\mbox {\tiny SS}}[\uprho]=
\int \left[\partial\sqrt{\uprho}/\partial z\right]^2 d^3 \bm r-
Z\int{\uprho(\bm r)\over \vert \bm r\vert}d^3\bm r\nonumber \\
+{1\over 2}\int\int 
{\uprho(\bm r)\uprho(\bm r')\over\vert\bm r-\bm r'\vert}\,d^3\bm r d^3\bm 
r'
\label{dm14}
\end{eqnarray}
with the subsidiary conditions
\begin{eqnarray}
&&\int\uprho(\bm r)\,d^3r\leq N\qquad \mbox{and}\nonumber \\
&&\int\uprho(\bm r)\,dz\leq B/(2\pi)\quad\mbox{for all } \bm 
r_\perp.\label{dm15}
\end{eqnarray}
In fact, for $\eta\geq\eta_c$, the minimizer of (\ref{dm:n_1}) has the form
\be \Gamma^{\mbox{\tiny DM}}_{\bm r_\perp}(z,z')=
\sqrt{\uprho^{\mbox{\tiny DM}}(\bm r_\perp,z)}
\sqrt{\uprho^{\mbox{\tiny DM}}(\bm r_\perp,z')},\ee
and (\ref{dm:n_1}) evaluated for $\Gamma^{\mbox{\tiny DM}}$ is the same as 
(\ref{dm14}) evaluated for $\uprho^{\mbox{\tiny DM}}$. Atoms in SS theory 
have the form of a thin {\em cylinder} with axis in direction of the 
magnetic field and with a cone-shaped region essentially cut out of its 
interior. The radius is finite, $R=\sqrt{2Z/B}$. The extension along the 
field is infinite, but the bulk of the electrons is confined within a 
distance 
$\sim Z^{-1}[\ln(B/Z^3)]^{-1}$ from the nucleus.

An even greater simplification occurs in the extreme limit 
$\eta\to\infty$, which we refer to as the {\em hyperstrong} (HS) case. In 
this limit the atom becomes effectively one dimensional and is described 
by a functional that can be minimized {\em in closed form}. This 
functional is
\be\E^{\mbox{\tiny HS}}[\rho]=\int\left[\partial\sqrt{\rho}/\partial 
z\right]^2dz-\rho(0)+
\int\rho(z)^2dz,\label{dm17}\ee
where $\rho(z)$ is a {\em one dimensional} density and the subsidiary 
condition is 
\be\int\rho(z)\,dz\leq \lambda=N/Z.\label{dm18}\ee
The connection between the SS and HS theories is as follows. 
Let $E^{\mbox{\tiny SS}}(N,B,Z)$ denote the minimum of (\ref{dm14}) with 
the subsidiary conditions (\ref{dm15}), and let $E^{\mbox{\tiny 
HS}}(\lambda)$ denote the minimum of
(\ref{dm17}) with the subsidiary condition (\ref{dm18}).
Let $L(\eta)$ be the solution to the equation
\be (\eta/2)^{1/2}=L(\eta)\sinh(L(\eta)/2).\label{dm:0-35}\ee
Then we have
\be E^{\mbox {\tiny SS}}(N,B,Z)=Z^3L(\eta)^2E^{\mbox{\tiny 
HS}}(\lambda)+Z^3O(L(\eta))\label{dm20}.\ee
There is also a corresponding connection between the minimizing densities,
$\uprho^{\mbox{\tiny SS}}_{N,B,Z}(\bm r)$ and $\uprho^{\mbox{\tiny 
HS}}_{\lambda}(z)$ for the two theories. Namely, 
\be[Z^2L(\eta)]^{-1}\int\uprho^{\mbox{\tiny SS}}_{N,B,Z}(\bm r_\perp,
[ZL(\eta)]^{-1}z)d^2\bm r_\perp\to \uprho^{\mbox{\tiny 
HS}}_{\lambda}(z)\label{dm21}\ee
(in the sense of distributions).

The function $L(\eta)$ behaves like $\ln \eta$ for large $\eta$, so the 
convergence of $E^{\mbox {\tiny SS}}$ to $E^{\mbox {\tiny HS}}$ is rather 
slow.
The main interest in the HS theory is that 
$\uprho^{\mbox{\tiny HS}}_{\lambda}(z)$ and
$E^{\mbox {\tiny HS}}(\lambda)$ can be explicitly computed: 
Writing $\uprho^{\mbox{\tiny HS}}_{\lambda}$ as $[\psi^{\mbox{\tiny 
HS}}_{\lambda}]^2$ one has
\begin{eqnarray}  \psi^{\mbox{\tiny HS}}_{\lambda}(z)&=&{\sqrt 
2(2-\lambda)\over4\sinh[\mfr1/4(2-\lambda)
\vert z\vert+c]}\quad\hbox{\hskip3pt for\ }\lambda<2\label{dm22}\\
 \psi^{\mbox{\tiny HS}}_{\lambda}(z)&=&\sqrt 2(2+\vert 
z\vert)^{-1}\quad\qquad\qquad
\hbox{for\ }\lambda\geq 2
\end{eqnarray}
where $\tanh c=(2-\lambda)/2$. Moreover\footnote{We recall that our energy 
unit is 54.4 eV},
\be E^{\HS}(\lambda)=-\mfr1/4\lambda+\mfr1/8\lambda^2
-\mfr1/{48}\lambda^3\ee
for $\lambda\leq 2$ and $E^{\HS}(\lambda)=E^{\HS}(2)=-1/6$ for 
$\lambda>2$. Thus  $\lambda_c=2$ in HS theory.
Eq.\ (\ref{dm20}) is essentially the statement of Theorem 3.5 in 
\cite{LSY94a}, but with one refinement: By replacing $\ln\eta$ in that 
theorem by $L(\eta)$ one obtains a neater estimate for the error term. The 
proof of (\ref{dm20}) and (\ref{dm21}), which follows closely the proof of 
Theorem 3.5 in \cite{LSY94a}, is given in the Appendix.

%%%%%%%%%%%%%%%%%%%%%%%%%%%%%%%%%%%%%%%%%%%%%%%%%%%%%%%%%%%%%%%%%%%%
\section{Numerical minimization of the density matrix functional}
In this section we describe in some detail the numerical methods used to 
study the DM theory. 
The task is to 
minimize numerically the density matrix functional (\ref{dm:n_1})
under the constraints (\ref{dm4}) and $N = \int\uprho_\Gamma (\bm 
r)\,d^3\bm r$.
The density matrix $\Gamma_{\bm r_\perp}$ is trace class and can be 
expressed in the form
\begin{equation}
\Gamma_{\bm r_\perp}(z,z') = \sum_{j=1}^\infty\lambda_j^{\bm r_\perp}
\phi_j^{\bm r_\perp}(z')^*\phi_j^{\bm r_\perp}(z)
\label{dm:n_2}
\end{equation}
where $\phi_j^{\bm r_\perp}(z)$ is an orthonormal basis in $L^2(\R,dz)$ for 
each $\bm r_\perp$
and $0\leq \lambda_j^{\bm r_\perp}\leq B/(2\pi)$, by condition 
(\ref{dm4}). For the minimizer $\Gamma^{\mbox{\tiny DM}}$ it turns out that
\begin{equation}
\lambda_j^{\bm r_\perp} =
\left\{
\begin{array}{cl}
{B/2\pi} & \mbox{if $j \leq j_c^{\bm r_\perp}$} \\
0 & \mbox{if $j > j_c^{\bm r_\perp}$}
\end{array}
\right.,
\label{dm:n_4}
\end{equation}
with $j \leq j_c^{\bm r_\perp}<\infty$. In fact,  $j \leq j_c^{\bm 
r_\perp}\leq j_c$ with a $j_c<\infty$ that is independent of $\bm r_\perp$ 
(but depends on $B$), cf. \cite{LSY94a}, p. 553. The functions 
$\phi_j^{\bm r_\perp}$ satisfy
the one dimensional Schr\"odinger equation
\begin{equation}
\left[
-{\partial^2\over\partial z^2} -{Z\over |\bm r|}+\int {\uprho_\Gamma (\bm 
r')\over
|\bm r-\bm r'|}\,d^3\bm r'
\right]\phi_j^{\bm r_\perp}(z) = \epsilon_j^{\bm r_\perp}\phi_j^{\bm 
r_\perp}(z)
\label{dm:n_5}
\end{equation}
and for each $(N,Z,B)$ there exists a 
unique $\mu^{\mbox{\tiny DM}}$ such that $j\leq j_c^{\bm r_\perp}$ if and 
only if $\epsilon_j^{\bm r_\perp}\leq\mu^{\mbox{\tiny DM}}$, and 
(\ref{dm:n_2}) is the solutions of the minimization problem.
Our strategy is to minimize the density matrix functional
(\ref{dm:n_1}) by iteratively solving the set of nonlinear eigenvalue 
equations
(\ref{dm:n_5}) and determining $\mu^{\mbox{\tiny DM}}$ such that the 
constraint
$\int\uprho_\Gamma=N$ is satisfied.
The eigenvalue equations are invariant with respect
to rotation around the $z$-axis. Hence they depend only on $|\bm 
r_\perp |$, but
even with this reduction (\ref{dm:n_5}) yields an infinite number of
eigenvalue equations, one for each value of $|\bm r_\perp |$.
We reduce them to a finite number 
by making the $|\bm r_\perp |$-axis discrete.
The DM-atom has a finite radius
$R\leq R_0=\sqrt{{2N/B}}$.
We therefore only have to consider the eigenvalue equations for which
$|\bm r_\perp |\leq R$. Let $N_{\bm r_\perp}$ be the number of eigenvalue 
equations
we choose to work with. Let 
\begin{equation}
\Delta_\perp = {R_0\over N_{\bm r_\perp} -1}.
\label{dm:n_8}
\end{equation}
We solve (\ref{dm:n_5}) at the $N_{\bm r_\perp}$ points $n\Delta_\perp$,
$n=1,..,N_{\bm r_\perp}$, on the $|\bm r_\perp |$-axis. Let
\begin{equation}
\theta^\perp_n (r ) =
\left\{
\begin{array}{ll}
1 & \mbox{if $r\in ((n-1)\Delta_\perp ,n\Delta_\perp ]$} \\
0 & \mbox{otherwise}
\end{array}
\right..
\label{dm:n_9}
\end{equation}
We minimize the density matrix functional (\ref{dm:n_1}) with
density matrices of the form
\begin{equation}
\Gamma_{\bm r_\perp}(z',z) =
\sum_{j=1}^\infty\sum_{n=1}^{N_{\bm r_\perp}}
\lambda_j^{n\Delta_\perp}
\phi_j^{n\Delta_\perp}(z')^*\phi_j^{n\Delta_\perp}(z)\theta_n^{\perp}(|\bm 
r_\perp |).
\label{dm:n_10}
\end{equation}
Let
\begin{equation}
h_n = -{\partial^2\over\partial z^2} -{Z\over\sqrt{(n\Delta_\perp )^2 
+z^2}},
\label{dm:n_11}
\end{equation}
and let $\hat\psi_i^n$ denote the eigenfunctions of $h_n$ and 
$\hat\mu_{i}^n$
the  corresponding
eigenvalues,
 so that $h_n\hat\psi_i^n = \hat\mu_{i}^n\hat\psi_i^n$.
We express $\phi_j^{n\Delta_\perp}(z)$ in terms of approximate
eigenfunctions
of $h_n$ which correspond to the $N_b$ lowest eigenvalues. Hence we write
\begin{equation}
\phi_j^{n\Delta_\perp}(z) = \sum_{i=1}^{N_b}c_{ji}^n\psi_i^n(z),
\label{dm:n_12}
\end{equation}
where $\psi_i^n(z)$ is an approximation for $\hat\psi_i^n$.
We determine approximations
for the basis functions $\hat\psi_i^n$ and their eigenvalues
$\hat\mu_i^n$ by the method of finite elements (FEM) for
eigenvalue problems,
dividing the interval $[-z_m,z_m]$, $z_m>0$ into elements\footnote{
The number of elements we use are $50-60$, and their width varies such
that the smallest elements are closest to the origin.}
and choosing a polynomial basis of degree $5$ within each, 
cf. \cite{FEM}. Let these solutions be $\psi^n_i(z)$ where
we induce the boundary condition that 
$\psi_i^n(\pm z_m) = 0$.
The eigenvalue corresponding to $\psi^n_i(z)$ is denoted by $\mu_i^n$.
Let $N_z$ be the number of samples of $\psi_i^n(z)$ values 
we choose to work with
along the $z$-axis and define
\begin{equation}
\Delta_z = {2z_m\over N_z-1}.
\label{dm:n_14}
\end{equation}
Then the samples we work with are $\psi_i^n ((l-1/2)\Delta_z -z_m)$,
$l = 1,\ldots ,N_z$.
We use the values $\mu_i^n$ as approximations
for $\hat\mu_i^n$. This basis is chosen because it is expected to be 
close to the solutions $\phi_j^{n\Delta_\perp}(z)$ and as such is a natural
starting point for the self-consistent iterations.
We have now
defined the set over which we numerically minimize (\ref{dm:n_1}).

We solve the set of nonlinear equations (\ref{dm:n_5}) in a self-consistent
manner iteratively.
 We define the chain of potentials
\begin{equation}
V_1^{(k)}(\bm r) = (1-\alpha)V_1^{(k-1)}(\bm r)+\alpha V_0^{(k-1)}(\bm r)
\label{dm:n_17}
\end{equation}
for $k>0$ and $V_1^{(0)} = 0$,  $\alpha \in (0,1]$\footnote{
Self-consistent iterations of the kind discussed here are in general not
convergent. The coefficient $(1-\alpha)$ acts as a
damping factor on the iterations.
The value of $\alpha$ is in practice chosen by trial and error
 as high as possible without inducing instability.
That way the iterations converge as fast as is possible.
}
. Let $\phi_j^{n\Delta_\perp,(k)}(z)$ be the
eigenfunctions of the operator
\begin{equation}
H^{n\Delta_\perp ,(k)} = h_n+V_1^{(k)}(\bm r)
\label{dm:n_18}
\end{equation}
with
\begin{equation}
V_0^{(k)}(\bm r) = \int {\Gamma_{x'_\perp}^{(k)}(z',z')\over |\bm r-\bm 
r'|}
\,d^3\bm r'
\label{dm:n_19}
\end{equation}
where
\begin{eqnarray}
\label{dm:n_20}
&&\Gamma_{\bm r_\perp}^{(k)}(z',z) = \\
&&\sum_{j=1}^{N_b}\sum_{n=1}^{N_{\bm r_\perp}}
\lambda_j^{n\Delta_\perp , (k)}\phi_j^{n\Delta_\perp , (k)}(z')^*
\phi_j^{n\Delta_\perp , (k)}(z)\theta_n^\perp (|\bm r_\perp |).
\nonumber
\end{eqnarray}
Note that $\phi_j^{n\Delta_\perp , (0)}(z) = \psi_j^n(z)$. With an 
appropriate
choice of $\alpha$ (we use $\alpha\in [0.01,0.1]$) the sequence
$\phi_j^{n\Delta_\perp , (k)}(z)$ turns out to be convergent and
\begin{equation}
\lim_{k\rightarrow\infty}\phi_j^{n\Delta_\perp , (k)}(z) =
\phi_j^{n\Delta_\perp }(z)
\label{dm:n_21}
\end{equation}
in $L^2([-z_m,z_m])$. This defines our self-consistent iteration
scheme. 
To be consistent with the discrete form of (\ref{dm:n_20}) we work
with the potentials $V_1^{(k)}$ and $V_0^{(k)}$ of the form
\begin{equation}
V(z,\bm r_\perp ) = \sum_{l=1}^{N_z}\sum_{n=1}^{N_{\bm r_\perp}}V_{ln}
\theta_l^\parallel (z)\theta^\perp_n (\bm r_\perp )
\end{equation}
where
\begin{equation}
\theta^\parallel_l (z ) =
\left\{
\begin{array}{ll}
1 & \mbox{if $z\in ((l-1)\Delta_z-z_m ,l\Delta_z-z_m ]$} \\
0 & \mbox{otherwise}
\end{array}
\right..
\label{dm:n_15}
\end{equation}
To determine (\ref{dm:n_19}) we calculate the boundary values on the
$N_{\bm r_\perp}\times N_z$ grid we work on by direct integration,
\begin{eqnarray}
&&\int {\Gamma_{x'_\perp}^{(k)}(z',z')\over |\bm r-\bm r'|}
\,d^3\bm r'
= \\ &&
\frac{2}{\sqrt{r}}\int_0^\infty\! dr'\sqrt{r'}
\Gamma_{x'_\perp}^{(k)}(z',z')
\mbox{Q}_{-\frac{1}{2}}
\left(
\frac{r^2+r'^2}{2rr'} \nonumber
\right)
,
\end{eqnarray}
where $\mbox{Q}_{\nu-\frac{1}{2}}$ is an associated Legendre function.
The direct integration of (\ref{dm:n_19}) is very slow but 
it is on the other hand
 very accurate. To determine $V_0^{(k)}$ at interior grid points faster
numerically we do the following:
We note that (\ref{dm:n_19}) is the solution of the Poisson equation
\begin{equation}
\nabla^2V_0^{(k)} = \Gamma_{\bm r_\perp}^{(k)}.
\label{dm:n_22}
\end{equation}
With the boundary values determined by the direct integration we use
standard five point difference approximation to the Laplacian 
in order to determine $V_0^{(k)}$ at the interior points.
We solve the finite difference scheme by simultaneous over-relaxation. We 
find that an over-relaxation coefficient of $1.8$ yields fast convergence
for this problem for our choice of $N_{\bm r_\perp}$ and $N_z$. 
%For an overview of the method of finite differences and
%simultaneous over-relaxation see eg. \cite{NR}. 
When we use the
finite difference scheme we predetermine
$V_0^{(k)}$ at a few interior grid points 
by direct integration. We then perform the over-relaxation
iterations until we obtain agreement with
 the predetermined values up to a desired
accuracy. We choose to iterate until the first $5$ digits of the solution
are identical to all the predetermined values.

Now we are ready to go through one iteration of the self-consistent
iteration scheme in detail. Let us consider the $k$-th step in the 
iteration.
At the start of this step we know
$\Gamma_{\bm r_\perp}^{(k-1)}$ (recall that
$\phi_j^{n\Delta_\perp,(0)} = \psi_j^n$). First we determine $V_0^{(k-1)}$ 
and then $V_1^{(k)}$ according to the scheme described above.
We next determine the matrix elements
\begin{eqnarray}
H^{n\Delta_\perp , (k)}_{lm} &=& \int \psi_l^nH^{n\Delta_\perp , 
(k)}\psi_m^n
\,dz \nonumber \\
&=& \delta_{lm}\tilde\epsilon_l^n + \int \psi_l^nV_1^{(k)}(\bm 
r)\psi_m^n\,dz
\label{dm:n_23}
\end{eqnarray}
for each $n = 1,\ldots ,N_\perp$. Now the eigenvectors of the matrices 
$H^{n\Delta_\perp , (k)}_{lm}$
correspond to the coefficients $c_{ji}^{n,(k)}$ in
\begin{equation}
\phi_j^{n\Delta_\perp , (k)}(z) = 
\sum_{i=1}^{N_b}c_{ji}^{n,(k)}\psi_i^n(z),
\label{dm:n_24}
\end{equation}
and we denote their corresponding eigenvalues by 
$\epsilon_j^{n\Delta_\perp , (k)}$.
I.e.,
\begin{equation}
H^{n\Delta_\perp , (k)}\phi_j^{n\Delta_\perp , (k)}(z) = 
\epsilon_j^{n\Delta_\perp , (k)}\phi_j^{n\Delta_\perp , (k)}(z).
\label{dm:n_25}
\end{equation}
To determine the eigenvectors and eigenvalues of the matrices 
$H^{n\Delta_\perp , (k)}_{lm}$ we use the eigen-routines from \cite{NR}.
Finally, at the end of the iteration step we determine $\mu^{\mbox{\tiny 
DM},(k)}$
 such that
$\int \Gamma_{\bm r_\perp}^{(k)}(z,z)\,d^3\bm r = N$. We are then ready 
for the
next step of the iteration.

We continue the iterations until ${\cal E}^{\mbox{\tiny 
DM}}(\Gamma^{(k)})$
\lq\lq stops\rq\rq\ changing.
More precisely we choose to stop when the change in
${\cal E}^{\mbox{\tiny DM}}(\Gamma^{(k)})$ between iterations
is in the $7$-th digit.

We
calculate with $N_\perp = 101$, $N_z = 201-301$ and $N_b = 30 - 60$.
We chose $z_m$ in such a way that $2z_m$ is at least three times longer
than the length of the atom the calculation yields. We obtained this
criterion by increasing $z_m$ until the ground state energy we obtained
became stable in the first $6$ digits.

To calculate the energy ${\cal E}^{\mbox{\tiny DM}}$ we note that
\begin{eqnarray}
E' &:=& \sum_j\int\lambda_j^{\bm r_\perp}\epsilon_j^{\bm r_\perp}\,d^2 \bm 
r_\perp
 \nonumber \\
&=& \sum_{j=1}^{N_b}\sum_{n=1}^{N_{\bm r_\perp}} 2\pi n \Delta_\perp^2
\lambda_j^{n\Delta_\perp}\epsilon_j^{n\Delta_\perp} \nonumber \\
&=& K^{\mbox{\tiny DM}} - A^{\mbox{\tiny DM}} +2R^{\mbox{\tiny DM}}.
\label{dm:n_26}
\end{eqnarray}
Here $K^{\mbox{\tiny DM}}$, $A^{\mbox{\tiny DM}}$ and $R^{\mbox{\tiny DM}}$
are  respectively the kinetic, attractive and repulsive parts of 
${\cal E}^{\mbox{\tiny DM}}$.
With our choice of basis $\psi_i^n$ and $V(\bm r)$ as the attractive
potential we have
\begin{eqnarray}
E'' &:=&
\int
\left[
-{\partial^2 \Gamma_{\bm r_\perp}(z,z')\over\partial z'^2}
+  V(\bm r)\Gamma_{\bm r_\perp}(z,z)
\right]_{z'=z}
\,d^3\bm r \nonumber \\
&=& \sum_{j=1}^{N_b}\int
\lambda_j^{\bm r_\perp}\phi_j^{\bm r_\perp}(z)
\left[
-{\partial^2 \over\partial z^2}
+  V(\bm r)\right]
\phi_j^{\bm r_\perp}(z) \,d^3\bm r \nonumber \\
&=&
\sum_{j=1}^{N_b}\sum_{l=1}^{N_b}\sum_{k=1}^{N_b}\sum_{n=1}^{N_{\bm 
r_\perp}} \nonumber \\
&&\int\lambda_j^{n\Delta_\perp}c_{jl}^nc_{jk}^n
\theta_n^\perp (|\bm r_\perp |)\psi_l^n(z)h_n\psi_k^n(z)
\,d^3\bm r \nonumber \\
&=&
\sum_{j=1}^{N_b}\sum_{l=1}^{N_b}\sum_{k=1}^{N_b}\sum_{n=1}^{N_{\bm 
r_\perp}}
2\pi n\Delta_\perp^2\lambda_j^{n\Delta_\perp}c_{jl}^nc_{jk}^n\delta_{lk}
\mu_l^n \nonumber \\
&=&
\sum_{j=1}^{N_b}\sum_{l=1}^{N_b}\sum_{n=1}^{N_{\bm r_\perp}}
2\pi n\Delta_\perp^2\lambda_j^{n\Delta_\perp}(c_{jl}^n)^2\mu_l^n
\nonumber \\
&=& K^{\mbox{\tiny DM}} - A^{\mbox{\tiny DM}}.
\label{dm:n_27}
\end{eqnarray}
Since
\begin{equation}
A^{\mbox{\tiny DM}} = - \int V(\bm r)\Gamma_{\bm r_\perp}(z,z)\,d^3\bm r
\label{dm:n_28}
\end{equation}
we obtain
\begin{eqnarray}
E^{\mbox{\tiny DM}} &=& {1\over 2}(E'+E'') \\
K^{\mbox{\tiny DM}} &=& E'' + A^{\mbox{\tiny DM}} \\
R^{\mbox{\tiny DM}} &=& {1\over 2}(E'-E'').
\end{eqnarray}
This is how we evaluate the ground state energy and the terms it is 
composed of.
Regarding the  accuracy 
in the evaluation of $K^{\mbox{\tiny DM}}$, one should be aware of the 
fact that $A^{\mbox{\tiny DM}}$
is in general a lot larger than $K^{\mbox{\tiny DM}}$. In the case of the
STF-theory $A^{\mbox{\tiny STF}} = 15K^{\mbox{\tiny STF}}$ which is the
low magnetic 
field strength limit for the DM-theory.
 Therefore a slight relative error 
in $A^{\mbox{\tiny DM}}$ yields a much larger relative error in 
$K^{\mbox{\tiny DM}}$. Based on the information given above, we
estimate the numerical error of the scheme to be about $\pm 1$ in the 
fourth
digit of the ground state energy. We therefore show the first $4$ digits
when we present our results of the energy.

The estimate of $\lambda_c$ is done in the following way. 
$E^{\rm DM}(\lambda,\eta)$
is calculated as a function of $\lambda$ which is a strictly convex 
function
in our approximation, due to the finite box. The minimum of this function 
then determines
$\lambda_c$. However, the function is extremely flat around the minimum so 
it is
difficult to determine its position. We make a linear approximation of 
$E^{\rm DM}(\lambda,\eta)$ using two close lying points below the minimum.
The value thus obtained gives a lower bound on $\lambda_c$ close to the 
true value.

%%%%%%%%%%%%%%%%%%%%%%%%%%%%%%%%%%%%%%%%%%%%%%%%
\section{Atomic properties in DM theory}

The results of our numerical computations are presented in Tables I-VI and 
illustrated in Figures 1-5. As remarked before, iron is of special 
importance in astrophysical context, and for this reason we state our 
results for the reference value $Z=26$. The scaling relations (\ref{dm8}) 
and (\ref{dm9}) allow an easy transformation to other values. Moreover, 
{\em in this section magnetic fields are measured in Gauss and energies in 
keV} to facilitate comparison with astrophysical data and other 
computations. To transform into the units in which the original 
Hamiltonian (\ref{dm1}) is written it should be kept in mind that there 
the energy unit is 54.4 eV  and the unit for magnetic field strength is 
$9.40\times 10^9$ G. The dimensionless parameter $\eta=B/Z^3$ has for 
$Z=26$ and $B=10^{12}$ G the value $6.053\times 10^{-3}$. This may seem 
small for Region 4, but as discussed in \cite{LSY94b}, the relevant 
semiclassical parameter is really $\eta^{1/5}$ and at the quoted values of 
$Z$ and $B$ we have $\eta^{1/5}
=0.36$.

Figures 1 and 2 show contour plots of the electronic densities of iron 
atoms  with $N=Z$ according to DM theory at four different field 
strengths, ranging from $10^{11}$ to $10^{14}$ Gauss. It is evident that 
in the weakest field the bulk of the electrons is spherically distributed 
around the nucleus. With increasing field strength the sphericity gets 
more and more distorted. The atom is composed of cylindrical shells that 
decrease in number as the field goes up. Between $10^{13}$ and $10^{14}$ 
Gauss a transition to a cylindrical shape with a
single shell takes place. The number of shells corresponds to the number 
of eigenfunctions of the one dimensional Schr\"odinger operators
$-\partial^2/\partial z^2+V^{\mbox{\tiny DM}}_{\bm r_\perp}(z)$ that 
contribute to the density matrix in the sum (\ref{dm:n_2}). The critical 
value, $\eta_c$, at which this number has dropped to one, was determined 
numerically for $\lambda=1$ to be $\eta_c=0.148$. This corresponds to 
$B=2.44\times 10^{13}$ G for $Z=26$.

In Table I the ground state energy (\ref{dm6}) of iron is shown as a 
function of the field strength for $B$ between $10^{10}$ and $10^{14}$ 
Gauss and for various 
values of the ratio $\lambda=N/Z$ of electron number to nuclear charge. A 
comparison of some of these values with results obtained by other methods 
is given in the next section.

Table II shows the results tor $\lambda_c$ as a function of $B$ in DM 
theory.
At the extremely strong field of $10^{18}$ Gauss, corresponding to 
$\eta^{1/5}=5.7$, one finds $\lambda_c=1.232$, which is still quite far 
from the HS value $\lambda_c=2$. However, compared with Thomas Fermi 
theories, where $\lambda_c$ is always 1 \cite{L82},
the negative ionization is noticeable even for the weaker fields. The value 
$\lambda_c=1.046$ for iron at $B=10^{13}$ G corresponds to an excess 
negative charge of 1.2 electrons. The binding energy of 
the excess charge in DM theory is shown in Table III.

As remarked at the end of Section III a precise determination of 
$\lambda_c$
is difficult and the values quoted should be regarded as lower bounds at 
the respective field strengths.

\section{Comparison with other theories}

As discussed in Section II, DM theory simplifies in the two limits, 
$\eta\to 0$ and $\eta\to\infty$, which are respectively described by the 
STF and HS theories. In order to study the rate of this convergence we 
have 
compared the ground state energies at $\lambda=1$ in these three models in 
Table
IV and plotted them in Fig.\ 3. It is remarkable how closely the STF ground
state energy approximates the DM energy even at fields as strong as 
$10^{15}$ Gauss,
while the electronic densities in DM theory deviate appreciably from the 
spherical shape of STF theory already at $10^{12}$ Gauss as seen in Fig.\ 
1.
Thus in this case at least,  the energy calculations are much less 
sensitive to the details of the model than density calculations. There is, 
however, another way of comparing the densities in STF and DM theories. In 
Fig. 4 we have plotted together 
the STF density and the {\em spherically averaged} DM density, and one 
sees that
they are quite close for the bulk of the electrons, up to fields of the 
order $10^{12}$ G.

It is apparent from Fig 3 that the DM energy values approach the HS values 
as the field goes up, but the convergence is very slow and the asymptotic 
regime has not yet been reached at the strongest fields considered for the 
DM computations. On the other hand, it is interesting how well the HS 
density (\ref{dm22}) fits the DM density integrated over the cross section of the atom as shown in Fig. 5. 

Finally, in Tables V and VI we compare the ground state energy $E$ 
computed in 
the DM and STF theories with values obtained by some other methods in the 
literature. 
The comparison is made for iron at $10^{12}$ G, since this case has been
considered in a number of sources. In table V the value for DM theory is 
compared with Hartree Fock (HF) \cite{KOONIN87}, density functional 
\cite{J85}, \cite{KM88}, restricted variational (RV) \cite{M84}, 
Thomas-Fermi-Dirac (TFD) \cite{SO84} and STF calculations \cite{C74}. In 
Table 5 the splitting of the ground state energy into its various parts 
(kinetic, attractive, repulsive, exchange) is compared for DM, STF and HF 
calculations.

\section{Conclusions}

We have carried out a numerical study of the density matrix model that 
describes exactly the quantum mechanical ground state of atoms in a 
homogeneous magnetic field in the asymptotic limit when the nuclear charge 
$Z$, the electron number $N$ and the magnetic field $B$ tend to $\infty$ 
with $N/Z$ fixed and $B/Z^{4/3}\to\infty$. The calculations demonstrate the
following features of heavy atoms in high magnetic fields as the field 
strength increases: A transition from an approximately spherical shape to a
highly elongated shape, accompanied by a decrease in ground state energy 
and increasing ability to bind excess electrons. We have also compared the 
DM model with the semiclassical Thomas-Fermi theory and the one dimensional
density functional theory that describe respectively its low  and high 
field limits. When $(B/Z^3)^{1/5}$ is of order unity these simpler theories
are numerically and conceptually wrong and the full DM theory should be 
used. The quantitative agreement of DM theory with Hartree Fock 
calculations is quite good in strong fields; for iron at $B=10^{12}$ G the 
difference in binding energies is less than 2 \%. The DM theory, however, 
is much simpler computationally than HF theory for large atoms and appears 
suitable as a starting point for more refined approximations and for the
study 
of molecular binding in strong fields. 
%%%%%%%%%%%%%%%%%%%%%%%%%
\appendix
\section*{}

We give here a proof of Eqs.\ (\ref{dm20}) and (\ref{dm21}). It consists 
essentially in an improvement of the estimates in Proposition 3.3 in 
\cite{LSY94a}.

With $L(\eta)$ defined by (\ref{dm:0-35}) we define for any density $\rho$ 
a rescaled density $\rho_\eta$ by
\begin{equation}
\rho (\bm r_\perp ,z) = 
Z^4\eta\,L(\eta )\rho_\eta (Z\eta^{1/2}\bm r_\perp ,Z\,L(\eta )z). 
\label{dm:0:40}
\end{equation}
We can then write the SS functional (\ref{dm14}) as
\begin{equation}
{\cal E}^{\mbox{\tiny SS}}[\rho ] = Z^3\,L(\eta )^2
{\cal E}_\eta^{\mbox{\tiny SS}}[\rho_\eta ],
\label{dm:0_41}
\end{equation}
where  ${\cal E}_\eta^{\mbox{\tiny SS}}$ is defined by
\begin{eqnarray}
\label{dm:0_42}
{\cal E}_\eta^{\mbox{\tiny SS}}[\rho_\eta ] &=&
\int\left({\partial\sqrt{\rho_\eta}\over\partial z}\right)^2\,d^3\bm r
-\int\rho_\eta (\bm r) V_\eta (\bm r )\,d^3\bm r \\
&& +\,{1\over 2}\int\int \rho_\eta (\bm r)V_\eta (\bm r - \bm r')\rho_\eta 
(\bm r')
\,d^3\bm r\,d^3\bm r'\nonumber
\end{eqnarray}
with the rescaled Coulomb potential
\begin{equation}
V_\eta (\bm r ) = L(\eta )^{-1}(z^2+\eta^{-1}\,L(\eta )^2\bm 
r_\perp^2)^{1/2}.
\label{dm:0_43}
\end{equation}
The functional ${\cal E}_\eta^{\mbox{\tiny SS}}$ has a
ground state energy
\begin{eqnarray}
E_\eta^{\mbox{\tiny SS}}(\lambda ) &=& \inf\{
{\cal E}_\eta^{\mbox{\tiny SS}}[\rho_\eta ]:
\rho_\eta\in{\cal C}^{\mbox{\tiny
SS}},\int\rho_\eta\le\lambda,\nonumber \\
&&
\int\rho_\eta (\bm r_\perp,z)dz\le 1
\}.
\label{dm:0_44}
\end{eqnarray}
and a corresponding minimizing density denoted by by 
$\rho_\eta^{\mbox{\tiny SS}}$. It is related to $\uprho^{\mbox \tiny{SS}}$ 
by the scaling (\ref{dm:0:40}). The energy
$E_\eta^{\mbox{\tiny SS}}$ is related to $E^{\mbox{\tiny SS}}$ by the 
scaling
(\ref{dm:0_41}),
\begin{equation}
{ E}^{\mbox{\tiny SS}}(N,Z,B) = Z^3\,L(\eta )^2
{ E}_\eta^{\mbox{\tiny SS}}(\lambda ).
\label{dm:0_45}
\end{equation}
The improvement
of proposition 3.3 in \cite{LSY94a} is stated in the following lemma:
\begin{lemma}
For any choice of $\lambda$ and $T$ there is a constant $C(\lambda ,T)$
such that
\begin{equation}
\left|\int\rho (\bm r_\perp ,0)\,d^2\bm r_\perp -\int V_\eta\rho
\right| \le C(\lambda ,T)\,L(\eta )^{-1}
\label{dm:0_46}
\end{equation}
holds, provided $\rho\geq 0$ satisfies $\int\rho\le\lambda$, 
$\rho (\bm r_\perp ,z) = 0$ for $|\bm r_\perp |>\sqrt{2}$, and
\[
T[\rho ] = \int\left({\partial\sqrt{\rho (\bm r_\perp ,z)}\over\partial z}
\right)^2\,d^3\bm r \le T.
\]
We can choose
$C(\lambda ,T) =\lambda + 
8\sqrt{2}\lambda^{1/4}T^{3/4}+8\sqrt{\pi}T^{1/2}\ln (\sqrt{2}+\sqrt{6})$.
\end{lemma}

{\it Proof :}
Following the proof of proposition 3.3 \cite{LSY94a} we write the
difference on the left side of (\ref{dm:0_46}) as $A_1+A_2+A_3$
with
\begin{equation}
A_1 = -\int_{|\bm r_\perp |\ge 1} V_\eta (\bm r)\rho (\bm r)\,d^3\bm r,
\label{dm:0_47}
\end{equation}
\begin{equation}
A_2 = \int_{|\bm r_\perp |\le 1}V_\eta (\bm r)[\rho (\bm r_\perp ,0) -\rho 
(\bm r)]
\,d^3\bm r
\label{dm:0_48}
\end{equation}
and
\begin{equation}
A_3 = \int \left[1-\int_{|\bm r_\perp |\le 1}V_\eta (\bm r )\,dz \right]
\rho (\bm r_\perp ,0)d^2\bm r_\perp.
\label{dm:0_49}
\end{equation}
With the same arguments as in \cite{LSY94a} we obtain
\begin{equation}
|A_1|\le {\lambda\over L(\eta )},
\label{dm:0_50}
\end{equation}
\begin{equation}
|A_2| \le {8\sqrt{2}\over L(\eta )}\lambda^{1/4}T^{3/4}
\label{dm:0_51}
\end{equation}
and\footnote{In \cite{LSY94a}, p.\ 541 there is a  power of $1/2$ missing 
on $T$ in the estimate for $A_3$.}
\begin{eqnarray}
\label{dm:0_52}
&&|A_3| = \\ &&\left|
\int\left[ {2\over L(\eta )}\sinh^{-1}(\eta^{1/2}/(L(\eta )|\bm r_\perp 
|)) -1
\right]\rho (\bm r_\perp ,0)\,d^2\bm r_\perp\right| \nonumber \\
&&\le 2T^{1/2}\sqrt{2\pi}\nonumber \\ && \times
\left\{
\int_0^{\sqrt{2}} \left[ {2\over L(\eta )}\sinh^{-1}(\eta^{1/2}/(L(\eta 
)r)) -1
\right]^2r\,dr
\right\}^{1/2}. \nonumber
\end{eqnarray}
Now we deviate from \cite{LSY94a}. Estimating the integral in 
(\ref{dm:0_52})
we obtain
\begin{eqnarray}
|A_3|&\le & 2\sqrt{2\pi}T^{1/2}2^{1/4}\nonumber \\ && \times
\sup_{0\le r\le\sqrt{2}}
\left|\left[{2\over L(\eta )}\sinh^{-1}(\eta^{1/2}/(L(\eta )r)) 
-1\right]r^{1/2}
\right|\nonumber \\
&\le & 8\sqrt{\pi}T^{1/2}\ln (\sqrt{2}+\sqrt{6})/L(\eta ).
\label{dm:0_53}
\end{eqnarray}
The last inequality comes from the following.
If $\alpha = \sqrt{\eta}/L(\eta )$, then
\begin{equation}
2\sinh^{-1}(\alpha /\sqrt{2})-L(\eta ) = 0
\label{dm:0_54}
\end{equation}
by the definition (\ref{dm:0-35}) of $L(\eta)$. Using (\ref{dm:0_54}) we 
get
\begin{eqnarray}
&&\sup_{0\le r\le\sqrt{2}}
\left|(2\sinh^{-1}(\alpha /r) -L(\eta ))r^{1/2}
\right| \nonumber \\
&=& \sup_{0\le r\le\sqrt{2}}\left|
\left[2\sinh^{-1}(\alpha /r) -2\sinh^{-1}(\alpha /\sqrt{2}) 
\right.\right.\nonumber \\
&&+\left.\left.
2\sinh^{-1}(\alpha /\sqrt{2}) -L(\eta )\right]r^{1/2}
\right| \nonumber \\
&=& \sup_{0\le r\le\sqrt{2}}\left|
\left[2\sinh^{-1}(\alpha /r) -2\sinh^{-1}(\alpha /\sqrt{2})\right]r^{1/2}
\right| \nonumber \\
&=& 2\sup_{0\le r\le\sqrt{2}}\left|
\left[\ln (\sqrt{2}/r) + \ln\left({\alpha+\sqrt{\alpha^2+r^2}\over
\alpha+\sqrt{\alpha^2+2}}\right)\right]r^{1/2}
\right| \nonumber \\
&\le & 2\sup_{0\le r\le\sqrt{2}}\left|\ln (\sqrt{2}/r)r^{1/2}\right|
\nonumber \\ &&
+ 2\sup_{0\le r\le\sqrt{2}}\left|\ln\left({\alpha+\sqrt{\alpha^2+r^2}\over
\alpha+\sqrt{\alpha^2+2}}\right)r^{1/2}
\right| \nonumber \\
&\le& 2^{5/4}\ln 2 +2^{5/4}\ln ((1+\sqrt{3})/\sqrt{2}) \nonumber \\
&= &
2^{5/4}\ln (\sqrt{2}+\sqrt{6}).
\label{dm:0_55}
\end{eqnarray}
This concludes the proof.
\medskip

In analogy with proposition 3.4 in \cite{LSY94a} the following lemma
is a corollary of Lemma A.1:
\begin{lemma}
For $\rho$ as in lemma 1 there is a constant $C'(\lambda ,T)$ such that
\begin{eqnarray}
&&\left|
\int\int\rho (\bm r)V_\eta (\bm r -\bm r')\rho (\bm r')\,d^3\bm r\,d^3\bm 
r'
-\int\bar\rho (z)^2\,dz\right| \nonumber \\
&\le& C'(\lambda ,T) \,L(\eta )^{-1}
\end{eqnarray}
where $\bar\rho (z) = \int\rho (\bm r_\perp ,z)\,d^2\bm r_\perp$.
\end{lemma}

The proof of (\ref{dm20}) and (\ref{dm21}) is now identical to the proof 
of theorem 
3.4 in \cite{LSY94a} with lemmas 1 and 2 in place of Propositions 3.3 and 
3.4.

In order to illustrate the
difference between the function $L(\eta)$ and the
approximation $\L(\eta)\approx \ln \eta$ used in \cite{LSY94a}, the ratio 
$L(\eta )/\ln\eta$ is plotted in Figure 6.
\medskip

{\bf Acknowledgements.} This work was supported by the Icelandic 
Science Foundation, the Foundation for Graduate Studies 
and the Research Fund of the University of Iceland. We thank Sven Th.\
Sigurdsson and Chris Pethick for helpful advice and valuable comments.

%%%%%%%%%%%%%%%%%%%%%%%%%%%%%%%%%%%%%%%%%%%%%%%%%%%%%%%%%%%%%%%%%%%%
%\include{../cite}

\newpage
%%%%%%%%%%%%%%%%%%% Figures come here %%%%%%%%%%%%%%%%%%%%%%
\epsfverbosetrue
\vbox{
\begin{figure}
% fig1.ps: width=397.48041pt, height=361.34583pt
\epsfxsize= 198.740205pt
\epsfysize= 180.672915pt
\mbox{
\epsfbox{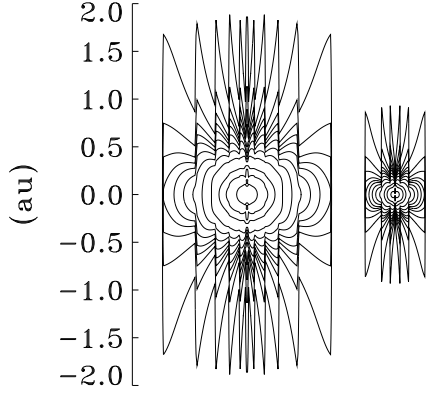}
}
\caption{Contour plots of the electronic density of iron atoms in DM 
theory for $B=10^{11}$ Gauss (left) and $B=10^{12}$ G (right). 
The outermost contour encloses 99 \% of the negative charge, the next  
90 \%, then 80 \% etc., and the two innermost 5 \% and 1\% respectively.}
\end{figure}
\vspace*{1.0cm}

\begin{figure}
\epsfxsize= 198.740205pt
\epsfysize= 180.672915pt
\mbox{
\epsfbox{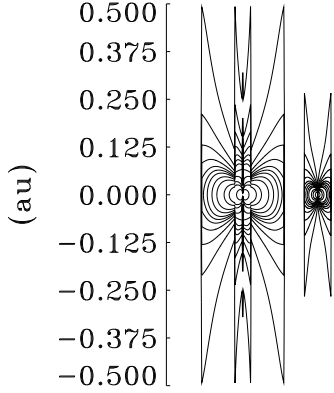}
}
\caption{Contour plots of the electron density of iron atoms in DM 
theory for $B=10^{13}$ Gauss (left) and $B=10^{14}$ G (right). The 
contours are drawn in the same way as in fig.\ 1. At 
$B=10^{14}$ the DM model has simplified and the density is described by 
the SS functional (12).}
\end{figure}
}

\vbox{
\begin{figure}
\epsfxsize= 198.740205pt
\epsfysize= 180.672915pt
\mbox{
\epsfbox{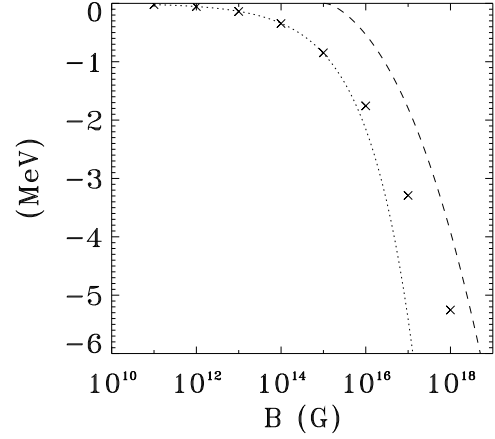}
}
\caption{The ground state energy of iron atoms as a function of the 
magnetic field strength $B$ in DM theory (crosses), STF theory (short 
dashes) and HS theory (long dashes).}
\end{figure}
\vspace*{1.7cm}

\begin{figure}
\epsfxsize= 198.740205pt
\epsfysize= 180.672915pt
%\begin{center}
\mbox{
\epsfbox{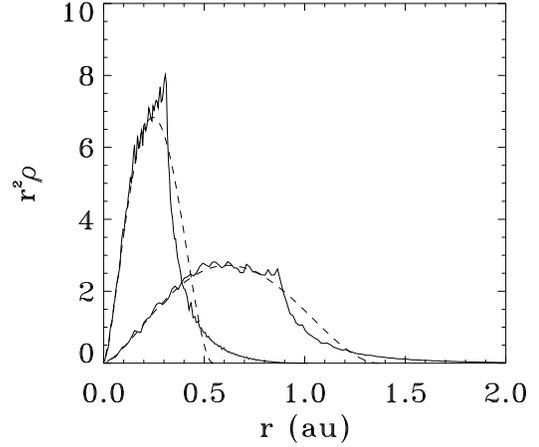}
}
%\end{center}
\caption{Comparison of the electron density in STF theory (dashed curve)
and 
the spherically averaged density in DM theory (solid curve) for 
$B=10^{11}$ Gauss (right) 
and $B=10^{12}$ G (left).}
\end{figure}
}

\begin{figure}
%fig5.ps: width=397.48041pt, height=568.11594pt
\epsfxsize= 198.740205pt
\epsfysize= 284.057970pt
%\begin{center}
\mbox{
\epsfbox{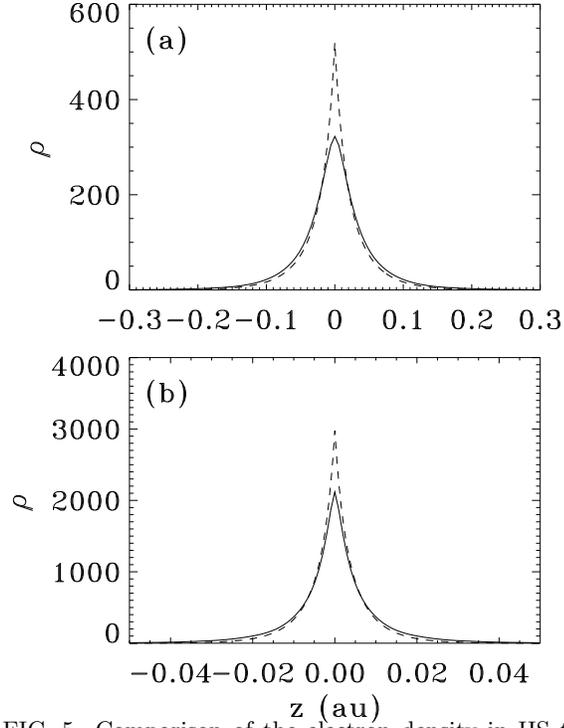}
}
%\end{center}
\caption{Comparison of the electron density in HS theory (dashed curve) 
and the integral over ${\bf r}_\perp$ of the density in DM theory (solid
curve) for $B=10^{13}$ Gauss (a) 
and $B=10^{18}$ G (b).}
\end{figure}

\begin{figure}
\epsfxsize= 198.740205pt
\epsfysize= 180.672915pt
\mbox{
\epsfbox{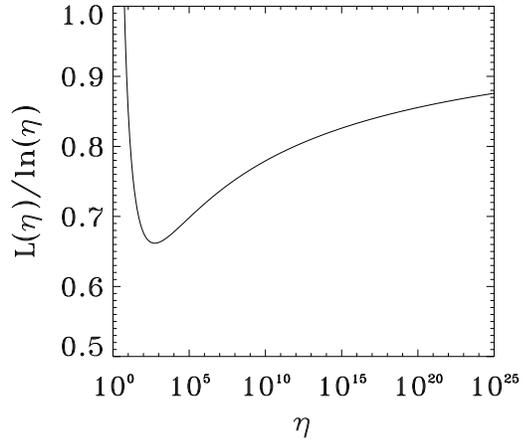}
}
\caption{The ratio $L(\eta)/\ln\eta$.}
\end{figure}

%%%%%%%%%%%%%%% Tables come here %%%%%%%%%%%%%%%%%%%%%%%%%

% Table 1
\newpage

\begin{table}
\caption{Ground state energy (in keV) of iron atoms (Z=26) as a function of
the 
magnetic field $B$ (in G) and the ratio $\lambda=N/Z$ of electron number 
to nuclear charge.}
\medskip
\begin{tabular}{dddddd}
$\lambda\backslash \,\,B$& $10^{10}$ & $10^{11}$ &$10^{12}$ &$10^{13}$ 
&$10^{14}$ \\ \hline 
0.1 & -3.202 & -8.188 & -20.80 & -52.58 & -122.4 \\
0.2 & -4.753 & -12.07 & -30.56 & -77.56 & -185.2 \\
0.3 & -5.838 & -14.80 & -37.47 & -95.04 & -230.2 \\
0.4 & -6.670 & -16.88 & -42.64 & -108.0 & -264.0 \\
0.5 & -7.277 & -18.41 & -46.56 & -117.9 & -289.4 \\
0.6 & -7.744 & -19.58 & -49.53 & -125.2 & -309.3 \\
0.7 & -8.102 & -20.42 & -51.73 & -130.6 & -323.7 \\
0.8 & -8.321 & -21.01 & -53.12 & -134.1 & -333.5 \\
0.9 & -8.475 & -21.41 & -53.85 & -136.6 & -339.9 \\
1.0 & -8.521 & -21.47 & -54.38 & -137.8 & -342.7 \\
\end{tabular}
\end{table}

% Table 2

\begin{table}
\caption{The ratio $\lambda_c=N_c/Z$ of the maximal negative charge to 
nuclear charge as a function of $B$ (in G) for iron (Z=26).}
\medskip
\begin{tabular}{cdcd}
$B$ & $\lambda_c$ & $B$ & $\lambda_c$ \\ \hline
$10^{10}$ & 1.020 & $10^{15}$ & 1.110 \\
$10^{11}$ & 1.026 & $10^{16}$ & 1.153 \\
$10^{12}$ & 1.035 & $10^{17}$ & 1.184 \\
$10^{13}$ & 1.043 & $10^{18}$ & 1.232 \\
$10^{14}$ & 1.061 &&\\
\end{tabular}
\end{table}
% Table 3

\begin{table}
\caption{The binding energy at maximal negative ionization 
in DM theory.}
\medskip
\begin{tabular}{cdcd}
$B$  & {${E^{{\rm DM}}(\lambda_c)-E^{\rm DM}(1) \over E^{\rm DM}(1)}$} &
$B$  & {${E^{{\rm DM}}(\lambda_c)-E^{\rm DM}(1) \over E^{\rm DM}(1)}$}\\
\hline
$10^{10}$ & 0.0011 & $10^{15}$ & 0.0073 \\
$10^{11}$ & 0.0008 & $10^{16}$ & 0.0123 \\
$10^{12}$ & 0.0035 & $10^{17}$ & 0.0232 \\
$10^{13}$ & 0.0038 & $10^{18}$ & 0.0203 \\
$10^{14}$ & 0.0050 & &  \\
\end{tabular}
\end{table}

% Table 3

%\begin{table}
%\caption{Comparison of the binding energy at maximal negative ionization 
%in DM theory with the difference between STF and DM theory.}
%\begin{tabular}{cdd}
%$B$  & {${E^{{\rm DM}}(\lambda_c)-E^{\rm DM}(1) \over E^{\rm DM}(1)}$} & 
%${E^{{\rm DM}}(1)-E^{\rm STF}(1) \over E^{\rm DM}(1)}$ \\ \hline
%$10^{10}$ & 0.0011 & -0.0057 \\
%$10^{11}$ & 0.0008 & -0.0028 \\
%$10^{12}$ & 0.0035 & 0.0057 \\
%$10^{13}$ & 0.0038 & 0.0145 \\
%$10^{14}$ & 0.0050 & 0.0044 \\
%$10^{15}$ & 0.0073 & -0.0972 \\
%$10^{16}$ & 0.0123 & -0.3266 \\ 
%$10^{17}$ & 0.0232 & -1.7629 \\ 
%$10^{18}$ & 0.0203 & -1.5803 \\ 
%\end{tabular}
%\end{table}

% Table 4

\begin{table}
\caption{Comparison of the ground state energy (in keV) of iron atoms in 
DM theory, STF theory and HF theory at various field strengths $B$ (in G). 
See also Fig.\ 3.}
\medskip
\begin{tabular}{cddd}
$B$  & $E^{\rm DM}$  & $E^{\rm STF}$ & $E^{\rm HS}$ \\ \hline
$10^{11}$ & -21.47 & -21.57 & \\
$10^{12}$ & -54.37 & -54.07 & \\
$10^{13}$ & -137.8 & -135.8 & \\
$10^{14}$ & -342.7 & -341.2 & \\
$10^{15}$ & -786.3 & -857.0 & -0.3346 \\
$10^{16}$ & -1623. & -2153. & -535.3 \\
$10^{17}$ & -3065. & -5405. & -1797. \\
$10^{18}$ & -5264. & -13583. & -3913.\\ 
\end{tabular}
\end{table}

% Table 5

\begin{table}
\caption{Ground state energy (in keV) of iron atoms at $B=10^{12}$ G
according to DM theory, HF theory [17], DF computations, denoted 
DF${}^a$ [15] and DF${}^b$ [16], RV computations [23], 
TFD theory [12], and STF theory [8].}
\medskip
\begin{tabular}{cddddddd}
 & DM & HF & ${\rm DF}^a$ & ${\rm DF}^b$ & RV & TFD & STF    \\ \hline
$E$ & -54.38 & -55.10 & -56.10 & -58.3 & -53.13 & -56.21 & -54.07     \\
\end{tabular}
\end{table}

% Table 6

\begin{table}
\caption{The composition of the ground state energy $E$ (in keV) of iron 
atoms at $B=10^{12}$ G in STF theory, HF theory and DM theory. 
$K$ is the kinetic energy, 
$A$ the attractive potential energy due to the nucleus, $R$ the 
energy of Coulomb repulsion and $E_{\rm ex}$ the exchange energy.}
\medskip
\begin{tabular}{cddddd}
& $E$  & $K$  &$A$  &$R$  &$E_{\mbox{\tiny ex}}$ 
\\
\hline
STF & -54.07&10.81&-97.33&32.44&0\\
HF &-55.10&10.6&-95.4&32.7&-3.06 \\
DM &-54.38&10.43&-96.90&32.09&0 \\
\end{tabular}
\end{table}
\end{multicols}
\end{document}